\documentclass[times,twocolumn,final]{elsarticle}

\usepackage{framed,multirow}

\usepackage{amssymb}
\usepackage{amsmath}
\usepackage{latexsym}

\usepackage{url}
\usepackage{xcolor}

\usepackage{hyperref}
\usepackage{cleveref}
\usepackage{algorithm}  
\usepackage{algpseudocode}

\usepackage{booktabs}
\definecolor{newcolor}{rgb}{.8,.349,.1}
\usepackage{geometry}
\begin{document}
	
	
	\begin{frontmatter}
		
		\title{Channel prior convolutional attention for medical image segmentation.}%
		
		\author[1]{Hejun Huang}
		
		\author[1,2]{Zuguo Chen\corref{cor1}}
		\cortext[cor1]{Corresponding author.}
		\ead{zg.chen@hnust.edu.cn}
		
		\author[1]{Ying Zou}
		\author[1]{Ming Lu}
		\author[1]{Chaoyang Chen}
		
		\address[1]{School of Information and Electrical Engineering, Hunan University of Science and Technology, Xiangtan 411201, China}
		\address[2]{Shenzhen Institute of Advanced Technology, Chinese Academy of Sciences, Shenzhen 518055, China}
		

		\begin{abstract}
			
			Characteristics such as low contrast and significant organ shape variations are often exhibited in medical images. The improvement of segmentation performance in medical imaging is limited by the generally insufficient adaptive capabilities of existing attention mechanisms. An efficient Channel Prior Convolutional Attention (CPCA) method is proposed in this paper, supporting the dynamic distribution of attention weights in both channel and spatial dimensions. Spatial relationships are effectively extracted while preserving the channel prior by employing a multi-scale depth-wise convolutional module. The ability to focus on informative channels and important regions is possessed by CPCA. A segmentation network called CPCANet for medical image segmentation is proposed based on CPCA. CPCANet is validated on two publicly available datasets. Improved segmentation performance is achieved by CPCANet while requiring fewer computational resources through comparisons with state-of-the-art algorithms.
			Our code is publicly available at \url{https://github.com/Cuthbert-Huang/CPCANet}.
			
		\end{abstract}
		
		\begin{keyword}
			
			\texttt Attention mechanism\sep Channel attention\sep Spatial attention\sep Medical image segmentation
		\end{keyword}
		
	\end{frontmatter}
	
	
	\section{Introduction}
	\label{sec1}
	Medical image segmentation is a crucial step in quantitatively assessing organ function \citep{Ramedsh_areviewmedical}. A vital role in medical image segmentation is played by deep learning methods \citep{Hesamian_Deeplearning}. The ability of convolutional neural network (CNN)-based approaches to capture global information and obtain accurate segmentation results is hampered by limited receptive fields. Self-attention mechanisms are leveraged by Transformer-based methods \citep{Vaswani_attentionisallyouneed} to effectively establish long-range feature dependencies. Nonetheless, their generalization capability on small datasets remains insufficient.
	
	Recent researches have focused on improving the aforementioned methods to address their limitations, and these improvements can be broadly categorized into two groups. The first group involves the incorporation of attention mechanisms into CNNs. Convolutional operations along the channel dimension of the feature maps in UNet \citep{Ronneberger_UNet} are embedded by Channel-UNet \citep{Chen_Channel-Unet} to capture spatial information mapping between pixels for liver and tumors segmentation. A spatial attention module is introduced by SA-UNet \citep{Guo_Sa-unet} between the encoder and decoder of the main network to enhance its representational capacity for retinal vessel segmentation. By integrating attention mechanisms into CNNs, these methods compensate for the limited receptive fields and improve network performance. The second group involves the combination of CNNs with Transformers. Transformer is introduced by TransBTS \citep{Wang_Transbts} into the bottleneck layers of a CNN encoder-decoder structure for brain tumors segmentation. The 3d medical input image is divided into fixed-size blocks by UNETR \citep{Hatamizade_UNETR} and directly fed into a Transformer encoder. Additionally, skip connections are incorporated to a CNN decoder for restoring the original input resolution. Self-attention is often employed as a spatial attention mechanism by this category of methods to capture global information. The combination of the powerful local information learning ability of CNNs with the long-range feature dependency modeling capability of Transformers unleashes the potential of both networks.
	
	In conclusion, attention mechanisms play a crucial role in medical image segmentation and computer vision as a whole. Attention mechanisms typically consist of channel attention and spatial attention components. "What" to pay attention to is determined by channel attention \citep{Guo_Attention-asurvey}. Channel attention was pioneered by \cite{Hu_SE}, who introduced SENet. The core of SENet is the squeeze-and-excitation (SE) block, which consists of two key steps: global average pooling and channel-wise weight calculation. The network is enabled to focus on important features by this block. "Where" to pay attention to is determined by spatial attention \citep{Guo_Attention-asurvey}. The spatial transformer module, proposed by \cite{Jaderberg_Spatialtransformer}, can learn translation, scaling, rotation, and other general distortions, allowing the network to focus on the most relevant spatial regions. The utilization of only one form of attention would result in a deficiency in focusing on one dimension. By combining both channel attention and spatial attention, the network gains the capability to adaptively select crucial objects and regions. The convolutional block attention module (CBAM), which sequentially combines channel attention and spatial attention, was introduced by \cite{Woo_cbam}. CBAM sequentially combines channel attention and spatial attention to enable the network to focus on informative channels and significant regions. However, the spatial attention map is computed by compressing channels, which leads to a consistent distribution of spatial attention weights for each channel during element-wise multiplication with the input features. This limitation restricts the adaptive capability of attention, as the spatial attention weights are not able to dynamically adjust based on the specific characteristics of each channel.
	
	\begin{figure*}[!t]
		\centering
		\includegraphics[scale=.55]{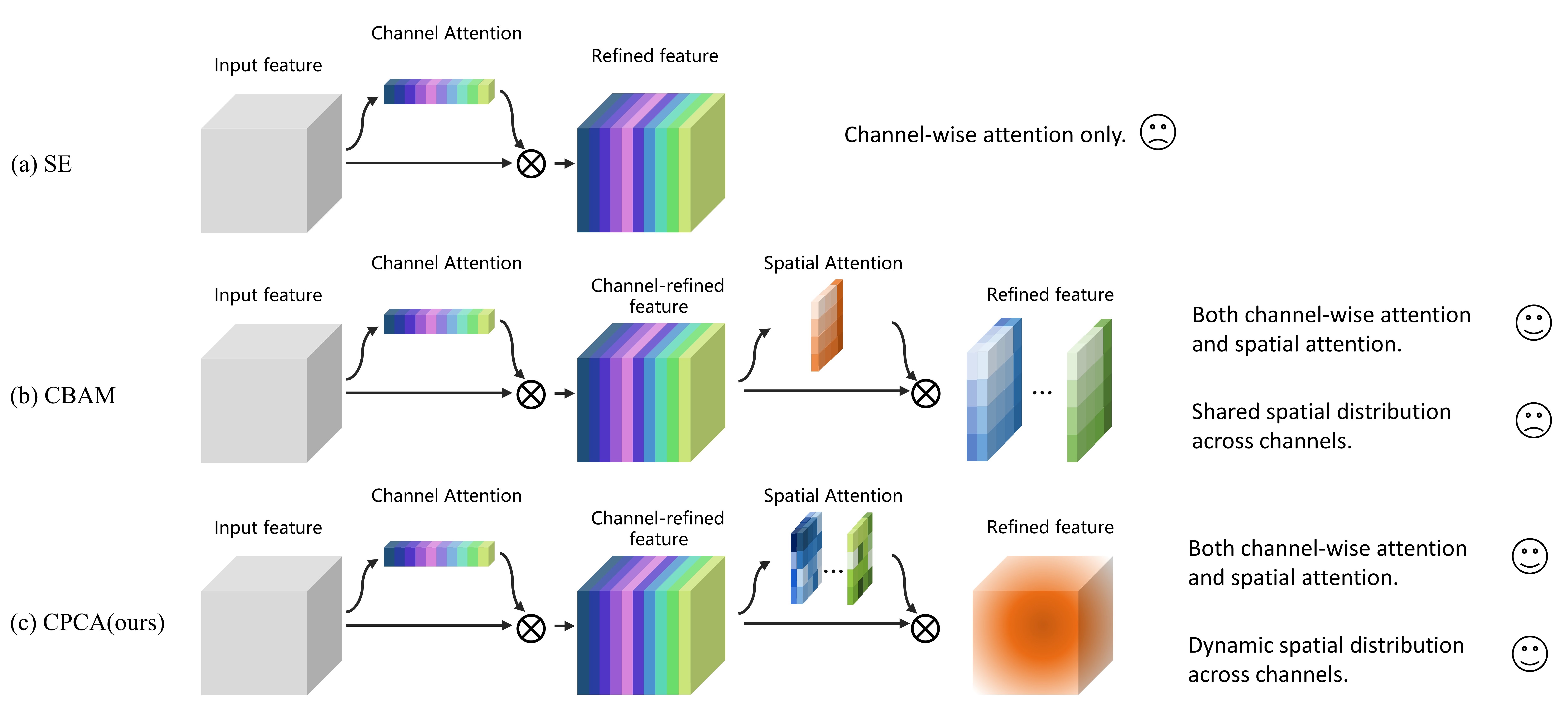}
		\caption{Schematic representation of the refined features of the three attention mechanisms (a) SE \citep{Hu_SE}, (b) CBAM \citep{Woo_cbam}, and (c) CPCA(ours).}
		\label{fig1}
	\end{figure*}
	
	This paper proposes a novel approach called Channel Prior Convolutional Attention (CPCA) that enables the dynamic distribution of attention weights in both the channel and spatial dimensions, as illustrated in Figure \ref{fig1} (c). Figure \ref{fig1} (b) shows that although CBAM incorporates both channel attention and spatial attention, it enforces a consistent spatial attention distribution across all channels in its output features. In contrast, SE (Figure \ref{fig1} (a)) only incorporates channel attention, which limits its ability to adequately select important regions. In contrast to CBAM, this study employs depth-wise convolutional modules to constitute the spatial attention component. The depth-wise convolution module employs strip convolution kernels of different scales to extract spatial mapping relationships between pixels. The use of multi-scale depth-wise strip convolution kernels ensures effective information extraction while reducing computational complexity \citep{Guo_SegNeXt,Huang_Ccnet}. The channel attention module is initially employed to obtain the channel attention map. Subsequently, the depth-wise convolution module sequentially extracts crucial spatial regions for each channel, resulting in dynamically distributed spatial attention maps on each channel. These dynamically distributed spatial attention maps on each channel closely approximate the actual feature distributions, effectively enhancing the network's segmentation performance. This convolutional attention mechanism is called Channel Prior Convolutional Attention (CPCA) and is based on prior channel attention.
	
	\begin{figure*}[!t]
		\centering
		\includegraphics[scale=.5]{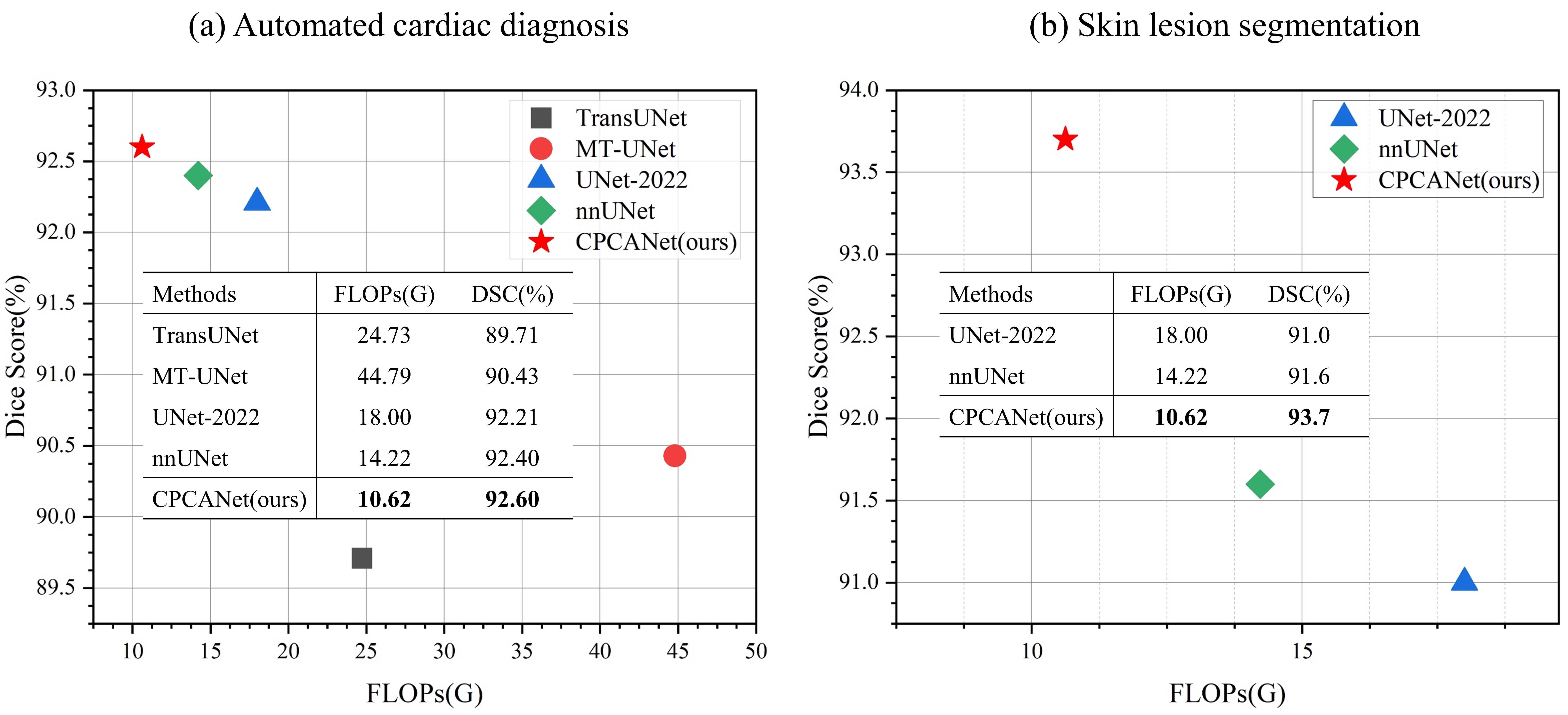}
		\caption{The evaluation of the proposed method with TransUNet \citep{Chen_Transttunet}, MT-UNet \citep{Jha_MT-UNET}, UNet-2022 \citep{Guo_UNet-2022}, nnUNet \citep{Isensee_nnUNet} methods for (a) Automated cardiac diagnosis \citep{Bernard_Deeplearning} and (b) Skin lesion segmentation \citep{Gutman_ISIC2016,Barata_Twosystems}. The computational complexity of each method is reflected in the FLOPs(G) (Floating Point Operations) metric, while the segmentation performance is measured by the DSC(\%) (Dice Similarity Coefficient).}
		\label{fig2}
	\end{figure*}
	
	The remarkable capacity for feature extraction forms the cornerstone of an exceptional network. CPCA is integrated into the backbone of Vision Transformer (ViT) \citep{Dosovitskiy_ViT,Han_asurvey-ViT} as a replacement for the self-attention mechanism. The backbone of ViT possesses an impressive capability for feature extraction, enabling the capture of global information. CPCA demonstrates effective attention towards valuable channels and significant regions. The combination of these two components manifests a powerful segmentation performance in our network, CPCANet, as demonstrated in Figure \ref{fig2}. Subfigure \ref{fig2} (a) presents the evaluation results for automated cardiac diagnosis \citep{Bernard_Deeplearning}, and subfigure \ref{fig2} (b) showcases the outcomes for skin lesion segmentation \citep{Gutman_ISIC2016,Barata_Twosystems}. The results indicate that, among the compared methods, CPCANet, proposed in this paper, achieves the best segmentation performance on both publicly available datasets while maintaining minimal computational complexity. Consequently, the contributions and novelty of this paper are summarised as follows:
	\begin{itemize}
		\item Utilizing the depth-wise convolution module to construct the spatial attention mechanism, which generates dynamically distributed spatial attention maps on each channel.
		\item Proposing a lightweight channel prior convolutional attention that generates attention maps closely approximating the real feature distribution.
		\item Introducing a segmentation network based on channel prior convolutional attention, which improves segmentation performance while reducing computational complexity.
	\end{itemize}
	
	\section{Related work}
	\label{sec2}
	\subsection{Medical image segmentation}
	\label{subsec2-1}
	
	CNN-based methods, such as U-Net \citep{Ronneberger_UNet}, have gained widespread popularity in medical image segmentation \citep{Siddique_unetreview}. Several variants of U-Net have been proposed and applied in this field. For retinal vessel segmentation, attention mechanisms and dense connections were incorporated into U-Net by \cite{Luo_AD-UNet}. Comprehensive skip connections and deep supervision based on U-Net were utilized by \citep{Huang_UNet3+} to effectively leverage multi-scale information. Attention gates were employed by MA-Unet \citep{Cai_Ma-unet} to address semantic ambiguity introduced by skip connections, and multi-scale prediction fusion was adopted to incorporate global information at different scales. However, despite their focus on attention embedding and multi-scale feature fusion in CNN-based methods, these studies were limited by the inability of CNNs to model long-range dependencies in features, which can significantly impact segmentation performance in challenging tasks.
	
	In recent years, Vision Transformers have gained significant attention for their capacity to model long-range dependencies. Nonetheless, Vision Transformers typically lack a robust local inductive bias and often necessitate training on large datasets to achieve desirable outcomes. Consequently, in the realm of medical image segmentation, many studies have resorted to combining CNNs with Vision Transformers. For example, TransUNet \citep{Chen_Transunet} integrated Transformers to extract contextual information from the feature maps generated by the CNN encoder, thereby enhancing the performance of medical image segmentation. Another approach, TransFuse \citep{Zhang_Transfuse}, employed a parallel combination of Transformers and CNNs to improve the efficiency of capturing global information.
	
	In this paper, a hybrid architecture is adopted for the proposed method. The self-attention of the Transformer layer is replaced by the Channel Prior Convolutional Attention (CPCA), which serves as a lightweight and efficient channel spatial attention mechanism. The modified Transformer layer is utilized in constructing the encoder, resulting in powerful feature attention capabilities and the ability to capture strong mapping relationships in deep features. To decode accurate segmentation results from the deep features with strong mapping relationships, a lightweight CNN is employed as the decoder, leveraging the strong inductive bias of CNN. Unlike previous approaches, the primary goal of this method is to achieve excellent segmentation performance while maintaining low computational complexity. The lightweight and efficient CPCA and decoder components play a crucial role in both achieving outstanding segmentation performance and reducing computational complexity.
	
	\subsection{Attention mechanism}
	\label{subsec2-2}
	Attention mechanisms serve the purpose of allowing networks to adaptively focus on important parts. For segmentation tasks, attention mechanisms can be categorized into channel attention and spatial attention. Important objects are the focal point of channel attention \citep{Chen_Channel-Unet,Hu_SE,Wang_ECA-Net}, whereas significant regions are the focal point of spatial attention \citep{Guo_Sa-unet,Jaderberg_Spatialtransformer}. Methods that combine Transformer primarily employs the self-attention mechanism of Transformers or enhances it. The Gated Axial-Attention model, introduced by MedT \citep{Valanarasu_Medicaltransformer}, extends existing architectures through the incorporation of supplementary control mechanisms within the self-attention module. A global spatial attention module, proposed by TransAttUnet \citep{Chen_Transttunet}, effectively assimilates long-range contextual interactions by combining Transformer self-attention with global spatial attention. Richer context dependencies are captured through the utilization of guided self-attention mechanisms, as proposed by \cite{Sinha_Muti-scaleself-guided}. PraNet \citep{Fan_PraNet} introduced reverse attention to enhance the precision of segmentation boundaries, yet it inadequately addresses significant regions. To overcome the computational burden associated with self-attention, SegNeXt \citep{Guo_SegNeXt} proposed a substitute in the form of a highly efficient multi-scale convolutional attention, yielding favorable outcomes. These techniques solely concentrate on regions of spatial significance, neglecting attention to vital objects within the channel dimension. CBAM \citep{Woo_cbam} amalgamates channel attention and spatial attention, but the spatial attention map is derived through channel compression, leading to consistent distribution of spatial attention weights across channels.
	
	A depth-wise convolutional module is used in this study to construct spatial attention, with individual calculation of the spatial attention map for each channel. The channel attention and spatial attention are combined to create the channel prior convolutional attention, facilitating the dynamic distribution of attention weights in both the channel and spatial dimensions. Multi-scale depth-wise stripe convolutional kernels are utilized by the depth-wise convolutional module, drawing inspiration from \cite{Guo_SegNeXt,Huang_Ccnet}, to significantly mitigate computational complexity. The proposed attention mechanism successfully enhances the adaptive capability while simultaneously ensuring lower computational complexity, thereby facilitating its straightforward integration into other networks.
	
	\section{Methods}
	\label{sec3}
	When formulating a novel attention module, two primary objectives are targeted:
	
	\textbf{Dynamic distribution of attention weights in both the channel and spatial dimensions.} Existing methods incorporate both channel attention and spatial attention but exhibit a uniform spatial attention weight distribution across channels. This consistent weight distribution introduces noise and severely impacts the network's segmentation performance.
	
	\textbf{Avoiding the significant computational overhead caused by complex operations.} Common self-attention mechanisms exhibit considerable computational complexity, rendering them computationally burdensome, particularly when handling large-resolution images and 3D images. Consequently, there exists an urgent demand for a lightweight yet high-performance attention mechanism capable of mitigating this computational burden.
	
	\begin{figure*}[!t]
		\centering
		\includegraphics[scale=.7]{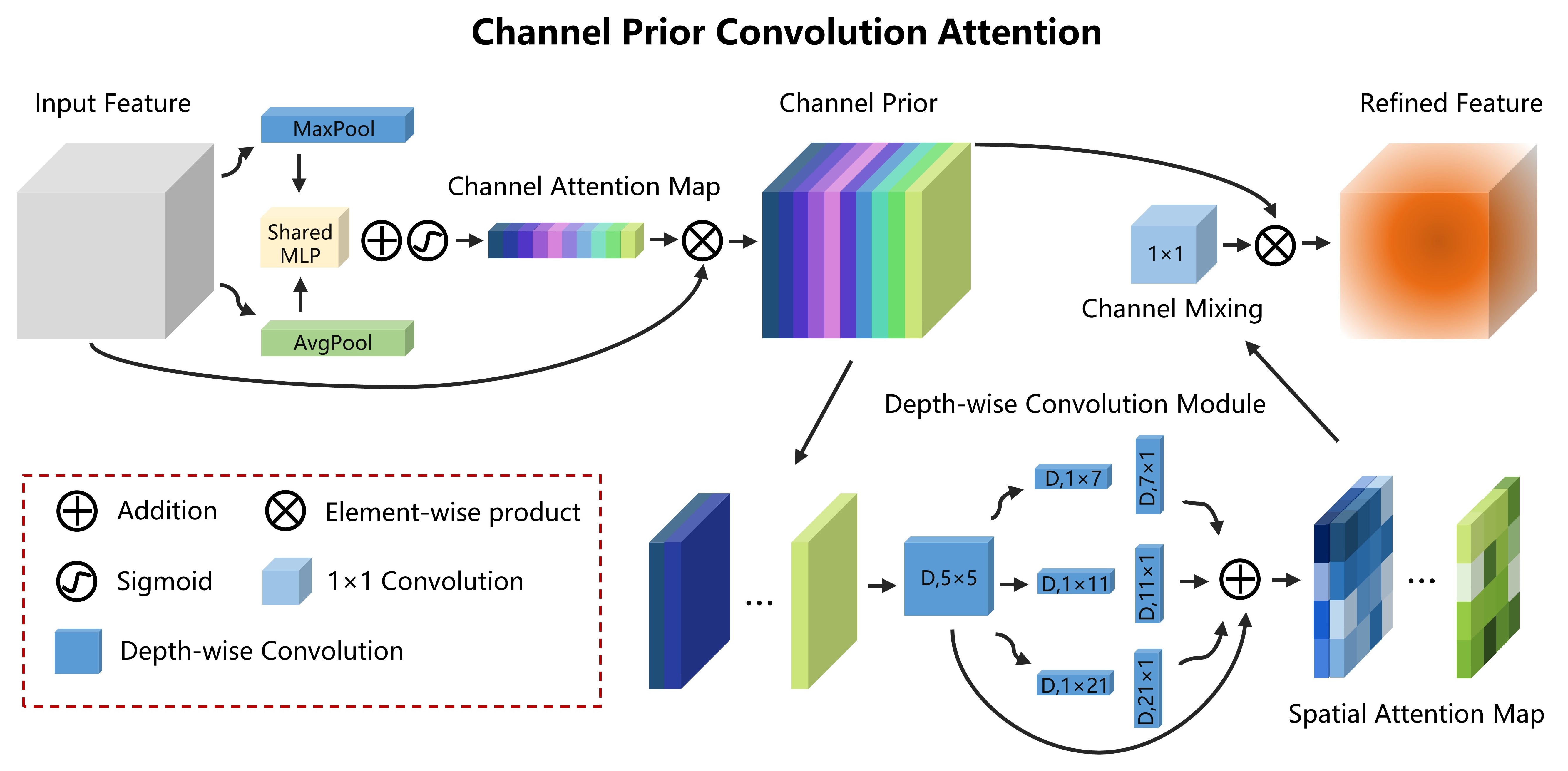}
		\caption{Channel Prior Convolutional Attention (CPCA) features an overall structure comprising sequential placement of channel attention and spatial attention. Spatial information of the feature maps is aggregated by the channel attention through operations such as average pooling and max pooling. The spatial information is subsequently processed through a shared MLP (Multi-Layer Perceptron) and added to produce the channel attention map. The channel prior is obtained by element-wise multiplication of the input feature and the channel attention map. Subsequently, the channel prior is inputted into the depth-wise convolution module to generate the spatial attention map. The convolutional module receives the spatial attention map for channel mixing. Ultimately, the refined features are obtained as the output by element-wise multiplication of the channel mixing result and the channel prior. The channel mixing process contributes to enhancing the representation of features.}
		\label{fig3}
	\end{figure*}
	
	\begin{figure*}[!t]
		\centering
		\includegraphics[scale=.6]{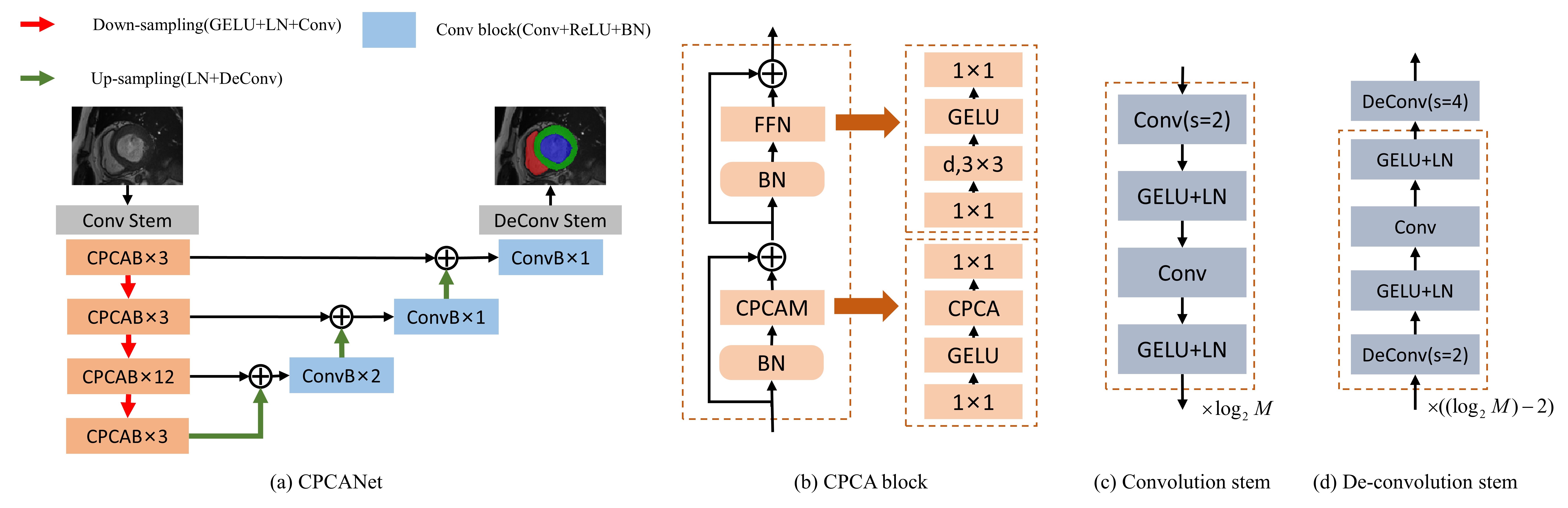}
		\caption{Illustrations of CPCANet. (a) illustrates the overall structure of CPCANet. (b) showcases the internal structure of the CPCA block. (c) and (d) depict the detailed structure of the convolution (Conv) and de-convolution (DeConv) stems. CPCAB, ConvB, LN, and BN represent the CPCA block, Conv block, layer normalization, and batch normalization, respectively. $M$ is a hyper-parameter, which varies based on the input resolution.}
		\label{fig4}
	\end{figure*}
	
	\subsection{Channel prior convolutional attention}
	\label{subsec3-1}
	The channel prior convolutional attention module performs channel attention and spatial attention sequentially, as shown in Figure 3. Given an intermediate feature map $F\in {{\mathbb{R}}^{C\times H\times W}}$ as input, the channel attention module (CA) first infers a 1D channel attention map ${{M}_{c}}\in {{\mathbb{R}}^{C\times 1\times 1}}$. ${{M}_{c}}$ is then element-wise multiplied with the input feature $F$, and the channel attention values are broadcast along the spatial dimension to obtain the refined feature ${{F}_{c}}\in {{\mathbb{R}}^{C\times H\times W}}$ with channel attention. The spatial attention module (SA) processes ${{F}_{c}}$ to generate a 3D spatial attention map ${{M}_{s}}\in {{\mathbb{R}}^{C\times H\times W}}$. The final output feature $\widehat{F}\in {{\mathbb{R}}^{C\times H\times W}}$ is obtained by element-wise multiplying ${{M}_{s}}$ with ${{F}_{c}}$. The overall attention process can be summarized as:
	\begin{equation}
		\begin{aligned}
			& {{F}_{c}}=\operatorname{CA}(F)\otimes F, \\ 
			& \widehat{F}=\operatorname{SA}({{F}_{c}})\otimes {{F}_{c}},
			\label{eq1}
		\end{aligned}
	\end{equation}
	where $\otimes $ represents element-wise multiplication.
	
	\textbf{Channel attention.} The generation of the channel attention map is the responsibility of the channel attention module, which achieves this by exploring the inter-channel relationships present within the features. Average pooling and max pooling operations are employed, following the approach proposed by CBAM \citep{Woo_cbam}, to aggregate spatial information from the feature map. This aggregation process yields two separate spatial context descriptors. These descriptors are subsequently fed into a shared multi-layer perceptron (MLP). The channel attention map is obtained by combining the outputs of the shared MLP through element-wise summation. In order to reduce parameter overhead, the shared MLP consists of a single hidden layer, where the size of the hidden activation is set to ${{\mathbb{R}}^{{}^{C}/{}_{r}\times 1\times 1}}$, with $r$ denoting the reduction ratio. The computation of the channel attention can be summarized as:
	\begin{equation}
		\begin{aligned}
			\operatorname{CA}(F)=\sigma (\operatorname{MLP}(\operatorname{AvgPool}(F))+\operatorname{MLP}(\operatorname{MaxPool}(F))),
			\label{eq2}
		\end{aligned}
	\end{equation}
	where $\sigma $ denotes the sigmoid function.
	
	\textbf{Spatial attention.} Spatial attention maps are generated through the extraction of spatial mapping relationships. In this study, it is believed that enforcing consistency in the spatial attention maps for each channel should be avoided. Dynamically distributing attention weights in both channel and spatial dimensions is considered to be more realistic. Figure 3 illustrates the utilization of depth-wise convolutions to capture spatial relationships among features, ensuring the preservation of inter-channel relationships while simultaneously reducing computational complexity. A multi-scale structure is employed to enhance the convolution operation's capability in capturing spatial relationships. Channel mixing is performed by the tail of the spatial attention module using a $1\times 1$ convolution, resulting in the generation of a more refined attention map. The computation of the spatial attention can be described as follows:
	
	\begin{equation}
		\begin{aligned}
			\operatorname{SA}(F)={{\operatorname{Conv}}_{1\times 1}}(\sum\limits_{i=0}^{3}{{{\operatorname{Branch}}_{i}}(\operatorname{DwConv}(F))}),
			\label{eq3}
		\end{aligned}
	\end{equation}
	where, $\operatorname{DwConv}$ represents depth-wise convolution, and ${{\operatorname{Branch}}_{i}},i\in \{0,1,2,3\}$ represents the $i$-th branch. ${{\operatorname{Branch}}_{0}}$ is the identity connection. Following the approach of \cite{Guo_SegNeXt,Huang_Ccnet}, we use two depth-wise stripe convolutions to approximate the standard per-channel convolution with a large kernel. The convolution kernel size for each channel is different to capture multi-scale information.
	
	Channel attention and spatial attention focus on "what" and "where" respectively and can be placed in parallel or in sequence \citep{Woo_cbam}. Experimental results have shown that sequential arrangement yields better results than parallel arrangement. In this paper, the design of spatial attention is based on the preceding order of channel attention, thereby forming channel prior convolutional attention.
	
	\subsection{Network overall architecture}
	\label{subsec3-2}
	A segmentation network based on channel-prior convolutional attention (CPCA) is proposed in this paper. The overall architecture is illustrated in Figure 4(a). A pyramid structure is adopted for the encoder, following previous works \cite{Guo_SegNeXt,Xie_SegFromer}. The encoder utilizes a CPCA block as its building block, which replaces the self-attention mechanism with CPCAM, as depicted in Figure 4(b). The CPCA block is compared with a Conv block (Conv+ReLU+BatchNorm (BN)) for the decoder, and experimental results indicate that the Conv block outperforms. This can be attributed to the strong inductive bias of CNN, which enables the decoding of precise segmentation results from deep features exhibiting strong mapping relationships.
	
	The encoder is composed of four stages with decreasing spatial resolutions, which correspond to four stages in the decoder with increasing spatial resolutions. Convolution stems and de-convolution stems with adjustable block numbers are employed to accommodate high-resolution input images, following the approach of UNET-2022 \citep{Guo_UNet-2022}. In the convolution stems, each block is comprised of two convolutional layers with strides of 2 and 1, respectively. Following each convolutional layer, an application of a GELU activation function and layer normalization is performed, as depicted in Figure 4(c). The number of blocks is determined by ${{\log }_{2}}M$, where $M$ represents the manually set downsampling factor. The de-convolution stem shares similar blocks with the convolution stems and also incorporates a transpose convolution with a stride of 4, as illustrated in Figure 4(d). The number of blocks is calculated as $({{\log }_{2}}M)-2$.
	
	\section{Experiments and results}
	\label{sec4}
	\begin{table*}[!t]
		\centering
		\caption{Quantitative comparison of multiple methods on automated cardiac diagnosis. The
			evaluation metrics are DSC (\%) and HD95 (mm). Params (M) represents the number of model parameters. FLOPs (G) represent floating point operations. The best results are indicated in bold.}
		\begin{tabular}{lcccccc}
			\toprule
			\multirow{2}[4]{*}{Methods} & \multicolumn{2}{c}{Average} & \multirow{2}[4]{*}{RV} & \multirow{2}[4]{*}{Myo} & \multirow{2}[4]{*}{LV} & \multirow{2}[4]{*}{FLOPs(G)} \\
			\cmidrule{2-3}          & DSC↑  & HD95↓ &       &       &       &  \\
			\midrule
			TransUNet \citep{Chen_Transunet} & 89.71 & -     & 88.86 & 84.54 & 95.73 & 24.73 \\
			SwinUNet \citep{Cao_Swin-unet} & 90.00  & -     & 88.55 & 85.62 & 95.83 & - \\
			MT-UNet \citep{Jha_MT-UNET} & 90.43 & -     & 86.64 & 89.04 & 95.62 & 44.79 \\
			MISSFormer \citep{Huang_Missformer} & 90.86 & -     & 89.55 & 88.04 & 95.83 & - \\
			UNet-2022 \citep{Guo_UNet-2022} & 92.21 & 2.555 & 90.07 & 90.49 & 96.05 & 18.00 \\
			nnUNet \citep{Isensee_nnUNet} & 92.40 & 1.225 & 90.67 & 90.40 & 96.14 & 14.22 \\
			\midrule
			CPCANet(ours) & \textbf{92.60} & \textbf{1.097} & \textbf{91.01} & \textbf{90.52} & \textbf{96.28} & \textbf{10.62} \\
			\bottomrule
		\end{tabular}%
		\label{tabl}%
	\end{table*}%
	
	\begin{figure*}[!t]
		\centering
		\includegraphics[scale=.9]{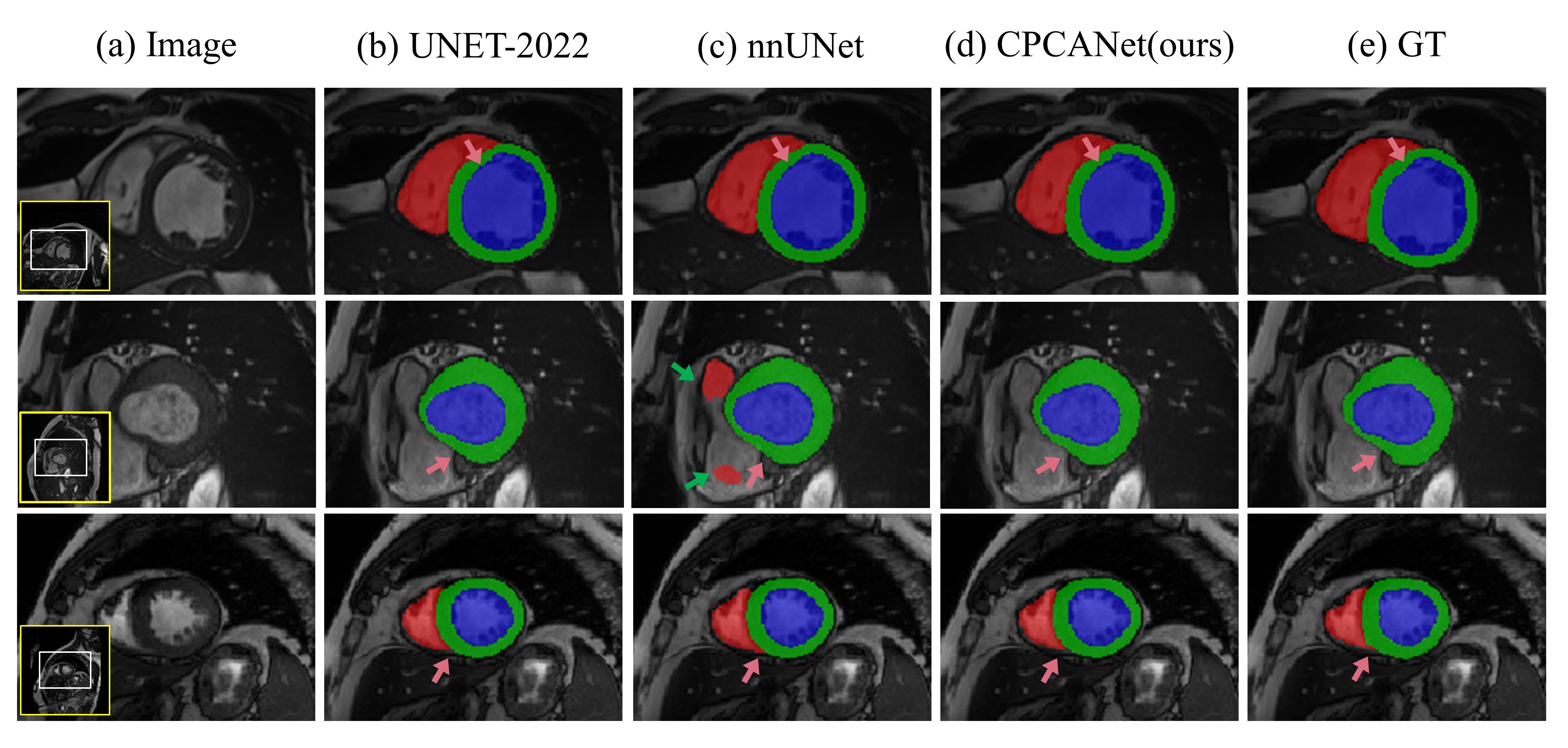}
		\caption{Visual comparison of the segmentation results on automated cardiac diagnosis. (a)Image, (b)UNET-2022 \citep{Guo_UNet-2022}, (c)nnUNet \citep{Isensee_nnUNet}, (d)CPCANet(ours), (e)GT. To enhance the visualization of the results, the presented image shows an enlarged region ((a) the white box in the bottom left corner). The color scheme is as follows: red represents the right ventricle (RV), green represents the myocardium of the left ventricle (MYO), and blue represents the left ventricle (LV).}
		\label{fig5}
	\end{figure*}
	
	\subsection{Dataset and pre-processing}
	\label{subsec4-1}
	Automated cardiac diagnosis (ACDC): The ACDC dataset \citep{Bernard_Deeplearning} consists of 100 samples, each containing images of the diastole and systole phases of the patient's heart. The right ventricle (RV), myocardium (MYO), and left ventricle (LV) need to be segmented. After preprocessing, the dataset yields a total of 1,902 slices. 70 samples are used for training (1,290 slices), 10 samples for validation (196 slices), and 20 samples for testing (416 slices). The evaluation metrics used are Dice Similarity Coefficient (DSC) and 95\% Hausdorff Distance (HD95).
	
	Skin lesion segmentation (ISIC-2016 and PH2): The ISIC-2016 dataset \citep{Gutman_ISIC2016} contains 900 samples for lesion segmentation from dermoscopic images. Following the approach in \cite{Guo_UNet-2022,Lee_Structureboundary,Wang_Boundary-awaretransformer}, the PH2 dataset \citep{Barata_Twosystems} is used as the test set, and the evaluation metrics employed are DSC and Intersection over Union (IoU).
	
	\subsection{Implementing details}
	\label{subsec4-2}
	The experiments in this paper were conducted on the PyTorch 1.12.0 + CUDA 11.3 framework using a NVIDIA RTX 3090 GPU. The code was implemented in Python 3.8.8. Both the entropy loss and dice loss were utilized, and they were combined using equation \ref{eq4}, where ${{\lambda }_{DC}}$ and ${{\lambda }_{CE}}$ were set to 1.2 and 0.8, respectively, for all datasets.
	
	\begin{equation}
		\begin{aligned}
			L={{\lambda }_{DC}}{{L}_{DC}}+{{\lambda }_{CE}}{{L}_{CE}},
			\label{eq4}
		\end{aligned}
	\end{equation}
	
	In the channel attention module, the reduction ratio was set to 16 to balance parameter overhead and modeling capacity. The embedding dimension was set to 96 for all datasets. For the ACDC dataset, a fixed crop size of $224\times 224$ was used, and $M$ was set to 4. For skin lesion segmentation, the images were cropped to a fixed size of $512\times 512$, and $M$ was set to 8. The crop size was chosen to be close to the original resolution of the datasets. The value of $M$ should increase as the crop size becomes larger to avoid a significant increase in computational cost when processing high-resolution images.
	
	Data augmentation was performed after random cropping to the fixed size. The specific augmentation methods followed the standard approaches used in nnUNet \citep{Isensee_nnUNet}. During inference, CPCANet made predictions using a sliding window approach. The stride for each sliding window was set to 0.5 times the crop size for both ACDC and skin lesion segmentation. A smaller stride means more overlapping patches are involved in the voting for mask prediction, resulting in improved segmentation performance. However, since the crop size for both ACDC and skin lesion segmentation datasets is close to the size of the entire image, reducing the stride does not lead to performance improvement but rather increases the inference time. Additionally, a Gaussian-weighted voting strategy was adopted for patch voting.
	
	\subsection{Performance on automated cardiac diagnosis}
	\label{subsec4-3}
	The performance and computational requirements of various methods in automated cardiac diagnosis are demonstrated through quantitative results and comparisons of computational complexity, as shown in Table \ref{tabl}. All the results are derived from a single model's accuracy, without the inclusion of ensembles, pre-training, or additional data. Average DSCs of 89.71\% and 92.40\% were achieved by TransUNet \citep{Chen_Transunet} and nnUNet \citep{Isensee_nnUNet}, respectively, among the existing works. However, they necessitated 24.73G and 14.22G FLOPs (floating-point operations). Conversely, the average DSC of the proposed CPCANet method improved to 92.60\%, and the required FLOPs were notably reduced to 10.62G. Additionally, CPCANet achieved an HD95 of merely 1.097mm, signifying a 10.4\% absolute error reduction in comparison to nnUNet.
	
	Visual segmentation results comparing the different methods are presented in Figure \ref{fig5}. While all three methods exhibit satisfactory segmentation performance, nnUNet, lacking attention mechanisms, displays notable missegmentation (highlighted by the green arrow in the second row of Figure \ref{fig5} (c)). Conversely, CPCANet, with its channel prior convolutional attention, effectively concentrates on crucial objects and regions, yielding superior segmentation details (indicated by the pink arrow in Figure \ref{fig5}).
	
	\subsection{Performance on skin lesion segmentation}
	\label{subsec4-4}
	\begin{table}[htb]
		\centering
		\caption{Quantitative comparison of multiple methods on skin lesion segmentation. The
			evaluation metrics are DSC (\%) and IoU (\%). The best results are indicated in bold.}
		\begin{tabular}{lcc}
			\toprule
			Methods & DSC↑  & IoU↑ \\
			\midrule
			SSLS \citep{Ahn_Automated} & 78.3  & 68.1 \\
			MSCA \citep{Bi_Automatedskin} & 81.5  & 72.3 \\
			FCN \citep{Long_Fullyconvolution} & 89.4  & 82.1 \\
			Bi et al. \citep{Bi_Dermoscopic} & 90.6  & 83.9 \\
			UNET-2022 \citep{Guo_UNet-2022} & 91.0  & 84.2 \\
			nnUNet \citep{Isensee_nnUNet} & 91.6  & 85.1 \\
			BAT \citep{Wang_Boundary-awaretransformer} & 92.1  & 85.8 \\
			\midrule
			CPCANet(ours) & \textbf{93.7} & \textbf{88.8} \\
			\bottomrule
		\end{tabular}%
		\label{tab2}%
	\end{table}%
	
	\begin{figure*}[!t]
		\centering
		\includegraphics[scale=.8]{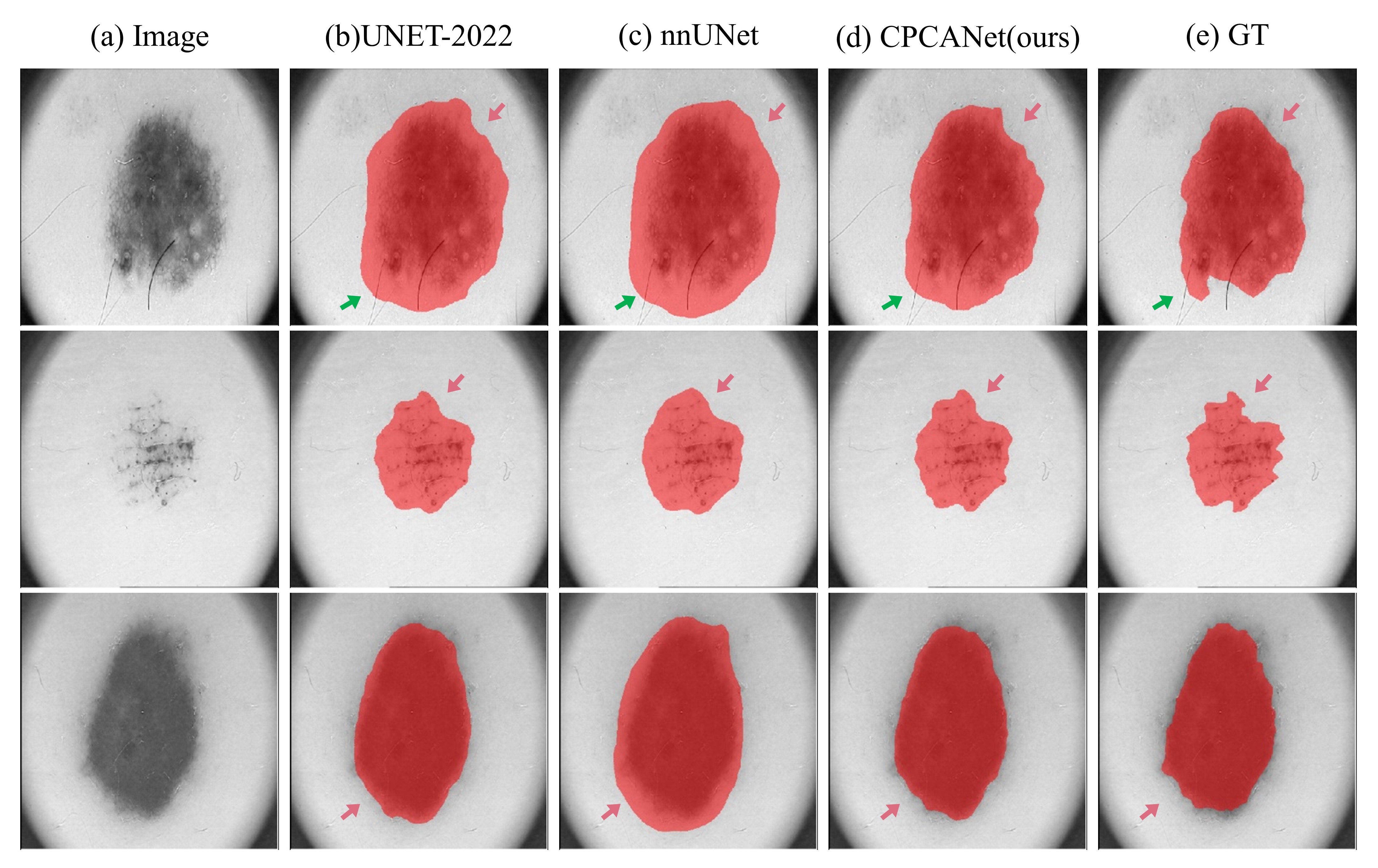}
		\caption{Visual comparison of the segmentation results on skin lesion segmentation. (a)Image, (b)UNET-2022 \citep{Guo_UNet-2022}, (c)nnUNet \citep{Isensee_nnUNet}, (d)CPCANet(ours), (e)GT.}
		\label{fig6}
	\end{figure*}
	Quantitative results comparing different methods for skin lesion segmentation are presented in Table \ref{tab2}. The DSC and IoU of multiple methods are reported. All the results are derived from a single model's accuracy, without the inclusion of ensembles, pre-training, or additional data. DSCs of 91.6\% and 92.1\% were achieved by nnUNet \citep{Isensee_nnUNet} and BAT \citep{Wang_Boundary-awaretransformer}, respectively, among the existing methods. Additionally, their IoUs were 85.1\% and 85.8\%. In comparison, the proposed CPCANet method demonstrates noteworthy enhancements over BAT. It achieves a DSC of 93.7\%, reflecting a 1.6 percentage points improvement, and an IoU of 88.8\%, reflecting a 3 percentage points improvement.
	
	Visual segmentation results comparing the different methods are presented in Figure \ref{fig6}. Challenging segmentation areas often include fine branches and local protrusions (highlighted by the green arrow in Figure \ref{fig6}). In these challenging areas, improved segmentation results are achieved by CPCANet, which has the capability to concentrate on important objects and spatial regions, compared to other methods. Moreover, when considering segmentation details (highlighted by the pink arrow in Figure \ref{fig6}), CPCANet generates results that closely align with the ground truth labels.
	
	\section{Discussion}
	\label{sec5}
	\begin{table*}[!t]
		\centering
		\caption{Ablation study of CPCA on the ACDC dataset. Channel-only means use channel attention only. Spatial-only means use spatial attention only. Parallel means that channel attention and spatial attention are placed in parallel, and Sequential means that the two attentions are placed according to sequential channel-spatial placement. The evaluation metrics are DSC (\%) and HD95 (mm). Params (M) represents the number of model parameters. FLOPs (G) represent floating point operations. The best results are indicated in bold.}
		\begin{tabular}{cccccccc}
			\toprule
			\multirow{2}[4]{*}{Methods} & \multicolumn{2}{c}{Average} & \multirow{2}[4]{*}{RV} & \multirow{2}[4]{*}{Myo} & \multirow{2}[4]{*}{LV} & \multirow{2}[4]{*}{Params(M)} & \multirow{2}[4]{*}{FLOPs(G)} \\
			\cmidrule{2-3}          & DSC↑  & HD95↓ &       &       &       &       &  \\
			\midrule
			Channel-only & 91.94 & 2.054 & 89.36 & 90.34 & 96.14 & \textbf{40.27} & 8.56 \\
			Spatial-only & 92.11 & 1.202 & 90.54 & 89.98 & 95.81 & 42.96 & \textbf{4.30} \\
			Parallel & 92.16 & 1.447 & 90.13 & 90.43 & 95.94 & 43.15 & 8.89 \\
			Sequential & \textbf{92.60} & \textbf{1.097} & \textbf{91.01} & \textbf{90.52} & \textbf{96.28} & 43.43 & 10.62 \\
			\bottomrule
		\end{tabular}%
		\label{tab3}%
	\end{table*}%
	\begin{table*}[!t]
		\centering
		\caption{Comparison of different spatial attention convolutional kernel sizes on the ACDC dataset. The three numbers in the brackets represent the kernel sizes of the three branches in the spatial attention module. The evaluation metrics are DSC (\%) and HD95 (mm). The best results are indicated in bold.}
		\begin{tabular}{lccccc}
			\toprule
			\multirow{2}[4]{*}{Kernel sizes} & \multicolumn{2}{c}{Average} & \multirow{2}[4]{*}{RV} & \multirow{2}[4]{*}{Myo} & \multirow{2}[4]{*}{LV} \\
			\cmidrule{2-3}          & DSC↑  & HD95↓ &       &       &  \\
			\midrule
			Kernels[3,5,7] & 91.86 & 2.429 & 89.78 & 89.99 & 95.81 \\
			Kernels[7,11,21] & \textbf{92.60} & \textbf{1.097} & \textbf{91.01} & \textbf{90.52} & \textbf{96.28} \\
			Kernels[11,21,41] & 92.41 & 1.122 & 90.72 & 90.45 & 96.06 \\
			\bottomrule
		\end{tabular}%
		\label{tab4}%
	\end{table*}%
	\begin{table*}[!t]
		\centering
		\caption{Comparison between CBAM and CPCA, as well as the comparison between CPCA with and without channel mixing, was conducted on the ACDC dataset. The evaluation metrics are DSC (\%) and HD95 (mm). The best results are indicated in bold.}
		\begin{tabular}{lccccc}
			\toprule
			\multirow{2}[4]{*}{Methods} & \multicolumn{2}{c}{Average} & \multirow{2}[4]{*}{RV} & \multirow{2}[4]{*}{Myo} & \multirow{2}[4]{*}{LV} \\
			\cmidrule{2-3}          & DSC↑  & HD95↓ &       &       &  \\
			\midrule
			CBAM  & 92.35 & 1.988 & 90.62 & 90.26 & 96.18 \\
			\midrule
			CPCA w/o channel mixing & 92.36 & 1.142 & 90.65 & 90.22 & 96.22 \\
			CPCA w/ channel mixing & \textbf{92.60} & \textbf{1.097} & \textbf{91.01} & \textbf{90.52} & \textbf{96.28} \\
			\bottomrule
		\end{tabular}%
		\label{tab5}%
	\end{table*}%
	\begin{table*}[!t]
		\centering
		\caption{Comparison of different decoders on the ACDC dataset. The numbers in square brackets represent the number of building blocks in each of the three stages of the decoder. The evaluation metrics are DSC (\%) and HD95 (mm). Params (M) represents the number of model parameters. FLOPs (G) represent floating point operations. The best results are indicated in bold.}
		\begin{tabular}{lccccccc}
			\toprule
			\multirow{2}[4]{*}{Decoder} & \multicolumn{2}{c}{Average} & \multirow{2}[4]{*}{RV} & \multirow{2}[4]{*}{Myo} & \multirow{2}[4]{*}{LV} & \multirow{2}[4]{*}{Params(M)} & \multirow{2}[4]{*}{FLOPs(G)} \\
			\cmidrule{2-3}          & DSC↑  & HD95↓ &       &       &       &       &  \\
			\midrule
			CPCABlock[1,1,1] & 92.00 & 1.364 & 89.86 & 90.04 & 96.11 & 43.35 & \textbf{8.96} \\
			CPCABlock[2,2,1] & 92.20 & 1.955 & 90.10 & 90.52 & 95.99 & 45.19 & 11.02 \\
			CPCABlock[3,3,2] & 92.31 & 1.120 & 90.72 & 90.20 & 96.00 & 47.14 & 12.05 \\
			ConvBlock[1,1,1] & 92.03 & 1.137 & 89.93 & 90.15 & 96.02 & \textbf{43.15} & 10.10 \\
			ConvBlock[2,2,1] & \textbf{92.60} & \textbf{1.097} & \textbf{91.01} & \textbf{90.52} & \textbf{96.28} & 43.43 & 10.62 \\
			ConvBlock[3,3,2] & 91.56 & 2.069 & 89.38 & 89.62 & 95.70 & 46.55 & 11.41 \\
			\bottomrule
		\end{tabular}%
		\label{tab6}%
	\end{table*}%
	\subsection{Attention ablation study}
	\label{subsec5-1}
	Ablation experiment were conducted on the ACDC dataset to validate the effectiveness and rationale of the proposed attention mechanisms. The performance of channel attention and spatial attention was evaluated individually, along with different ordering schemes. The results of the ablation experiment for CPCA on the ACDC dataset are presented in Table \ref{tab3}.
	
	It can be observed from the table that using either channel attention or spatial attention alone results in lower segmentation performance compared to their combined use. This verifies that better segmentation performance can be achieved by simultaneously focusing on informative objects and important regions. Moreover, the sequential channel-spatial ordering scheme demonstrates better performance compared to the parallel use of both attention modules. This finding validates the rationale behind the design of channel prior convolutional attention in this paper.
	
	\subsection{Effects of spatial attention kernel size}
	\label{subsec5-2}
	The feature extraction capability of the convolutional block is influenced by the size of the convolutional kernel. Global information can be effectively extracted by larger convolutional kernels due to their larger receptive field, thereby enhancing their ability to capture semantic information \citep{Peng_Largekernel}. Conversely, smaller convolutional kernels are effective in extracting texture information. Both a grasp of global information and the extraction of pixel-level details are required for semantic segmentation tasks. Hence, the selection of an appropriate convolutional kernel size is crucial to ensure the effectiveness of spatial attention. A comparison of different spatial attention convolutional kernel sizes on the ACDC dataset is presented in Table \ref{tab4}. The results suggest that convolutional kernel size combinations such as [7, 11, 21] can effectively leverage the benefits of spatial attention.
	
	\subsection{Effects of channel mixing}
	\label{subsec5-3}
	The final step of the proposed spatial attention in this paper involves channel mixing, which aims to further enhance feature representation. A comparison between CBAM and CPCA, as well as between CPCA with and without channel mixing, is presented in Table \ref{tab5}. Similar performance to CBAM in terms of DSC is observed when CPCA is used without channel mixing, but it exhibits a noticeable decrease in HD95. This suggests that CPCA has a significantly higher ability to focus on important regions compared to CBAM. Furthermore, the incorporation of channel mixing in CPCA results in a significant improvement in DSC.This confirms that channel mixing is effective in further enhancing feature representation.
	
	\subsection{Choice of decoder}
	\label{subsec5-4}
	The effectiveness of CPCABlock and ConvBlock in constructing the decoder was compared to create a more efficient decoder. The comparative results of different decoders on the ACDC dataset are presented in Table \ref{tab6}. It is demonstrated by the results that ConvBlock can achieve good segmentation performance even with fewer building blocks. Therefore, ConvBlock was chosen to construct the decoder, with the decoding process divided into three stages, each utilizing 2, 2, and 1 building blocks respectively.
	
	\subsection{Limitations and future work}
	\label{subsec5-5}
	Despite effectively focusing on useful objects and important regions, the proposed Channel Prior Convolutional Attention (CPCA) in this paper has limitations in achieving accurate segmentation boundaries, thereby restricting further improvement in segmentation performance. Furthermore, the segmentation network proposed in this paper is limited to a single size, making it unable to adapt to variations in dataset sizes. Future work will focus on enhancing the attention mechanism to achieve accurate segmentation boundaries and proposing segmentation networks that can adapt to multiple sizes. Additionally, the proposed method will undergo validation using more diverse and comprehensive datasets.
	
	\section{Conclusion}
	\label{sec6}
	Channel Prior Convolutional Attention (CPCA) is the proposed method in this paper, enabling dynamic distribution of attention weights in both channel and spatial dimensions. A depth-wise convolutional module is utilized to extract spatial relationships without impacting the channel prior. Multiple branches are employed in the depth-wise convolutional module to capture multi-scale information for fusion in the spatial attention map, thereby enhancing the overall spatial attention. Following the generation of the spatial attention map, channel mixing is conducted to further enhance feature representation. A segmentation network is designed based on CPCA and validated on two publicly available datasets. In comparison to existing methods, the proposed approach achieves enhanced segmentation performance with reduced computational resource requirements.
	
	\section*{Acknowledgments}
	The research has been supported by the Natural Science Foundation of Hunan Province - Youth Project (Project No. 2020JJ5201), the Natural Science Foundation of China (Project No. 21B0456) and the Natural Science Foundation of Shenzhen City (Project No. JCYJ20210324101215039).
	\bibliographystyle{model2-names.bst}\biboptions{authoryear}
	\bibliography{refs}
	
\end{document}